\documentclass[sigconf]{acmart}
\graphicspath{{Figures/}}
\usepackage{todonotes}
\usepackage{xspace}
\usepackage{wrapfig}
\usepackage{array}

\AtBeginDocument{%
  \providecommand\BibTeX{{%
    \normalfont B\kern-0.5em{\scshape i\kern-0.25em b}\kern-0.8em\TeX}}}

\copyrightyear{2024} 
\acmYear{2024} 
\setcopyright{acmlicensed}\acmConference[CHI '24]{Proceedings of the 2024 CHI Conference on Human Factors in Computing Systems}{May 11--16 2024}{Hawaii, USA}
\acmBooktitle{Proceedings of the 2024 CHI Conference on Human Factors in Computing Systems (CHI '24), May 11--16, 2024, Hawaii, USA}
\acmPrice{15.00}
\acmDOI{10.1145/3613904.3641971}

\begin{document}

\title{Towards A Diffractive Analysis of Prompt-Based Generative AI}

\author{Nina Rajcic}
\orcid{0000-0001-6501-5754}
\affiliation{%
  \institution{SensiLab, Monash University}
  \streetaddress{900 Dandenong Road}
  \city{Caulfield East}
  \state{Victoria}
  \country{Australia}
  \postcode{3145}
}
\email{Nina.Rajcic@monash.edu}
\author{Maria Teresa Llano}
\orcid{0000-0002-4898-1755}
\affiliation{%
  \institution{SensiLab, Monash University}
  \streetaddress{900 Dandenong Road}
  \city{Caulfield East}
  \state{Victoria}
  \country{Australia}
  \postcode{3145}}
\email{teresa.llano@monash.edu}

\author{Jon McCormack}
\orcid{0000-0001-6328-5064}
\affiliation{%
  \institution{SensiLab, Monash University}
  \streetaddress{900 Dandenong Road}
  \city{Caulfield East}
  \state{Victoria}
  \country{Australia}
  \postcode{3145}}
\email{Jon.McCormack@monash.edu}
\renewcommand{\shortauthors}{Rajcic, Llano and McCormack}

\begin{abstract}
Recent developments in prompt-based generative AI has given rise to discourse surrounding the perceived ethical concerns, economic implications, and consequences for the future of cultural production. As generative imagery becomes pervasive in mainstream society, dominated primarily by emerging industry leaders, we encourage that the role of the CHI community be one of inquiry; to investigate the numerous ways in which generative AI has 
the potential to, and already is, augmenting human creativity. In this paper, we conducted a diffractive analysis exploring the potential role of prompt-based interfaces in artists' creative practice. Over a two week period, seven visual artists were given access to a personalised instance of Stable Diffusion, fine-tuned on a dataset of their work. In the following diffractive analysis, we identified two dominant modes adopted by participants, AI for ideation, and AI for production. We furthermore present a number of ethical design considerations for the future development of generative AI interfaces.
\end{abstract}

\ccsdesc[300]{Applied computing~Fine arts}
\ccsdesc[500]{Computing methodologies~Artificial intelligence}
\ccsdesc[500]{Human-centered computing~Empirical studies in interaction design}
\keywords{Generative AI, Creative AI, Diffusion, Diffractive Analysis}

\begin{teaserfigure}
  \includegraphics[width=\textwidth]{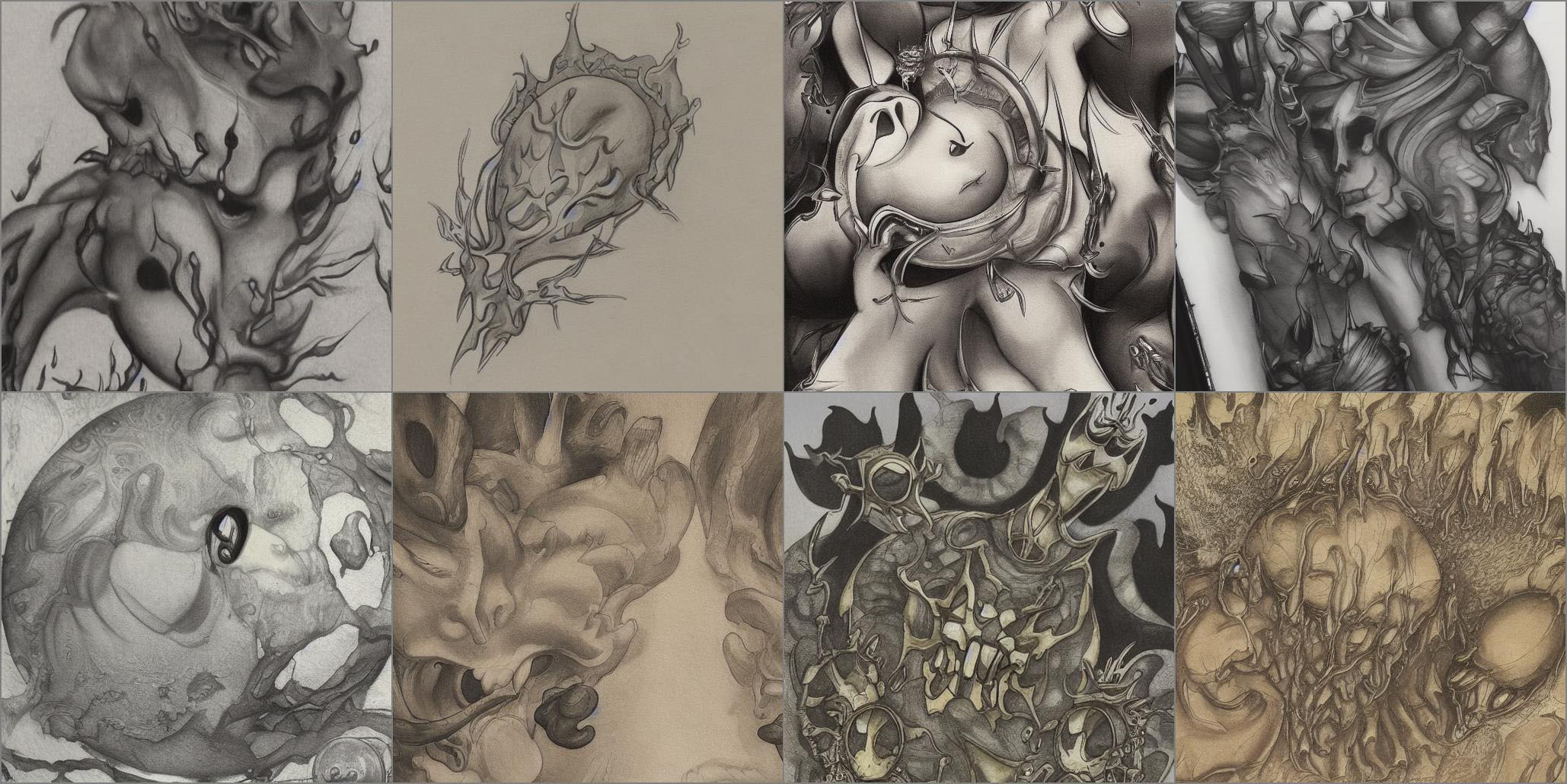}
  \caption{A selection of unprompted training samples output from finetuning Stable Diffusion on a set of visual artist's artwork}
  \label{fig:teaser}
\end{teaserfigure}

\maketitle

\section{Introduction}
\label{s:introduction}

Creative applications of generative Artificial Intelligence (AI) systems have garnered increasing interest in both research and industry, with systems such as DALL-E 2, MidJourney and Stable Diffusion allowing for the generation of detailed and complex imagery from short text descriptions. Despite the enormous technical feat undertaken, Text-to-Image (TTI) systems raise a number of ethical and moral dilemmas, giving rise to debate around copyright and plagiarism \cite{gross2023creative,samuelson2023generative}, artistic merit \cite{mikalonyte2022can} and economic implications \cite{hatzius2023potentially,ponce2023generative}.

Creative professionals are increasingly expressing their concern about the threat of AI replacing human illustrators and artists through the outsourcing of creative production. This concern is exacerbated by commercial makers of TTI systems often (over)claiming the artistic value of these systems' output. For instance, in the first release of Stable Diffusion it was claimed that it ``will empower billions of people to create stunning art within seconds''\footnote{\url{https://stability.ai/blog/stable-diffusion-announcement}}. However, the implications of these models, for diverse groups of people for which these tools are pitched to, are yet to be understood, leading practitioners in both research \cite{muller2022genaichi, kenthapadi2023generative} and industry to emphasise the importance of developing clear and rigorous guidelines towards the building and implementation of generative AI technologies. 

In this paper, we conduct a qualitative study to understand how visual artists conceptualise and integrate these AI tools. Rather than attempting to develop or contribute to technological innovation and quality of generative models, we instead push against the technological positivist approach of unbridled 'making', turning our focus towards the analysis and critique of the technologies that have already pervaded the space of generative AI. In this paper, we investigate how generative AI tools might be adopted by artists with an established creative practice founded in physical materials, as opposed to artists working primarily with TTI systems \cite{chang:2023}. We examine how the unique agency of AI models influence and challenges the role of the human creator, as well as the impacts on culture more broadly. In a time when industry-based and corporate research dominates AI development, we echo Fischer's proposition \cite{fischer2023generative} that one crucial role of academic research is to offer critical analysis of such already ubiquitous technologies, as well as build upon emerging methodologies in HCI \cite{rajcic2023message} that are designed to unpack increasingly pervasive and disruptive AI technologies.

Drawing inspiration from Karen Barad's notion of diffractive reading \cite{barad2018diffracting}, we demonstrate and build upon a new methodology in the space of Human Computer Interaction (HCI), Diffractive Methodology (DM). A diffractive analysis offers a nuanced and exploratory approach to evaluation, focusing not only on similarities, but on the differences across participant experiences, and ultimately tracing the potential source of these differences to produce deeper insight. In our analysis, we set out not to answer whether TTI systems are `good' or `bad' for creativity, rather to unpack the increasingly ambivalent attitudes towards the intrusion of the machine into realms historically reserved for the human.

Our work sheds light into the understanding of how visual artists, whose practice is based primarily in physical materials, may adopt fast developing AI technologies. Emerging from our diffractive analysis are two dominant modes of use: creative ideation and creative production. The contributions of this paper are realised as a set of {\em design considerations} specific to each mode (detailed in Section \ref{ss:design_considerations}).

\section{Background and Related Work}

\subsection{AI and Creativity} 
\label{sec:aicreativity}

The design of Creativity Support Tools (CST) has been identified as an important challenge for HCI research in order to advance, or enhance, people's creativity through technology \cite{shneiderman:2000,fischer:2004, chung2022artist}. In this area, efforts have been focused on tools that support the creative process by automating so called `non-creative' tasks (i.e. tasks that do not require problem solving, e.g. removing background from images). However, researchers in the field are increasingly turning their attention to the development of tools at different stages of the creativity process, arguing their potential to encourage divergent and convergent thinking \cite{frich2019mapping}. 

The rise of generative AI systems have moreover sparked debate about the nature of creativity itself, some argue that AI-generated art lacks the depth of human emotion and lived experience \cite{mccormack2023writing}, while others contend that creativity is not solely reserved to human actors \cite{colton2020machine}. The emergence of AI-generated artworks that evoke strong emotional responses challenges the notion that creativity is an exclusively human domain. 
Ultimately, generative AI technologies exhibit creativity largely due to their unpredictable nature; in their generative potential they possess the ability to surprise us, offering new visual and conceptual configurations, synthesising historical data in potentially novel forms. In this way, AI systems enact their own kind of unique creative agency.

\subsubsection{Co-Creative AI}
\label{sec:cocreative}

Researchers have mapped out the role of generative AI in the creative process \cite{inie2023designing, muller2022genaichi}, including Human-AI creative collaboration \cite{hwang2022too}, and generative imagery for creative ideation \cite{paananen2023using} \cite{chiou2023designing}. Numerous efforts have also been made to conceptualise models of human-machine collaborations in creative domains, some of these calling for a higher level of awareness from computational agents \cite{davis2015enactive, llano2022explainable}, while others have focused on real-time engagement and improvised interactions by enabling bidirectional channels of communication \cite{McCormack2019, thelle2021spire}. Mixed-initiative approaches have also been explored, in which both human and computational agents can initiate content exchanges through a mutual exploration of the design space \cite{yannakakis2014mixed, zhu2018explainable, lin2023beyond}. 

TTI systems are a new technology that offer a high level of flexibility and generality, yet to be comprehensively explored in the space of human-machine collaboration. Existing literature has predominantly focused on understanding the affordances of the text-based interface mostly through prompt engineering guidelines \cite{liu2022design,Ibarrola2023prompt}. A recent study of Stable Diffusion identified users with an artistic background to represent the second largest group in the community using TTI systems (17\% of the population in the study’s sample), with technical practitioners identified as the largest representative group (38\%) \cite{sanchez:2023}. This suggests that the limitations of TTI interfaces may prove prohibitive for non-technical artists. 
Prompting also overlooks creative activities that do not consist in the generation of final outputs, particularly related to the process, with artists reporting ``losing touch’’ with the physical world, rendering the creative process an intellectual task, emphasising the separation of mind and body otherwise intertwined in traditional artistic practices \cite{choiai:iccc23}. Capabilities such as model customization, exploration of the the design space, multi-modal input, and artists’ target prompt engineering tools, have been identified by \cite{Ko2023} as possible ways to enhance the adoption of prompting as a creative tool in artistic practice. Moreover, research has been done into perceptions of authorship in human-AI collaborations \cite{gero:2023, yuan:2022}. Our study contributes further understanding of the opportunities and challenges face by non-technical artists when integrating TTI systems.

Industry has largely outpaced and overshadowed the nuanced and varied research dedicated to the design and application of co-creative AI technologies. The entire field is dominated almost exclusively by one form of interaction: prompt-based generative AI. This raises crucial issues which we aim to address in this paper. Does a prompt-based interaction model fully encapsulate the preferred modes of collaboration between artists and generative AI? And what are the broader implications of this type of interface on the discourse on creativity and creative production?

\section{Diffractive Methodology}
\label{sec:diffractive}
In contemporary philosophical thought, posthumanism proposes a radical shift away from anthropocentrism, and a return to materiality \cite{braidotti2019posthuman}. Donna Haraway's exploration of the cyborg \cite{haraway2006cyborg} serves as an early precursor to this line of thinking, challenging the prevailing notions of secular humanism that elevate humans as the sole locus of agency, morality, rationality, and individuality. As advancements in AI continue to pervade and alter daily life, we also are witnessing the very definition of the 'human' become strained. 

Inspired by Barad's notion of diffractive reading, many have come to adopt a diffractive methodology as a qualitative method. Utilising the metaphor of the photon, Barad proposes that diffraction leads to a ``mapping of the effects of differences'', as opposed to reflection in which ``much like the infinite play of images between two facing mirrors, the epistemological gets bounced back and forth'' \cite{barad2003posthumanist}. Diffraction as a methodology is likewise proposed by Frauenberger \cite{frauenberger2019entanglement} as a valuable tool in HCI research---rather than asking what makes an artefact `work', we should instead ask how an artefact becomes `different things'.

Diffraction methodologies have been adopted in a number of fields across the social sciences, including sociology \cite{fox2023applied}, education research \cite{mazzei2014beyond}, and information systems \cite{osterlund2020building}. In \cite{osterlund2020building}, researchers employ and build upon Barad's photon metaphor in understanding the distinction between traditional analyses, and DM
\begin{quote}
``where refraction and reflection bracket the nature of light, diffraction can be used to study both the nature of light and the source of the light. It can tell you about an object and its traces at the same time''.
\end{quote}
 
The application of diffractive methodologies in HCI has received increased interest in recent years \cite{10.1145/3462326, nordmoen2022making}. In a recent study, researchers conduct a diffractive analysis to unpack the manifold implications of a domestic AI text-based system designed to augment memory \cite{rajcic2023message},. The diversity in approaches to diffractive analysis demonstrates that it does not encompass a fixed methodology, but rather an open reading of data `through' or alongside other materials such as ``texts, personal experiences, other data'' \cite{doi:10.1177/13607804211029978}.

In our case, DM is chosen due to the nature of the phenomena under enquiry. In nuanced matters such as art, creativity, identity, and agency, we argue that traditional methodologies fall short in capturing the full complexity of such entanglements between human and machine. This is due to their tendency to aggregate and simplify data for the purposes of reporting. Of course, every research methodology carries with it a particular lens which inevitably has a hand in shaping the presented results. For example, a Thematic Analysis (TA) will ``summarise qualitative data by artificially reducing its complexity and aggregating disparate events together''\cite{fox2023applied}. The objective of DM is to make explicit these methodological influences by tracing their effects. This approach departs from traditional methodologies (e.g. Thematic Analysis, Grounded Theory (GT)) firstly in its explicit conceptualisation of the research-assemblage as materially constitutive of the research findings. For instance, TA imposes the researchers' categorical structure onto data, and GT privileges data that neatly fits within an emergent structure \cite{fox2015inside}. Secondly, in drawing attention to differences, contradictions, and divergences, a diffractive analysis foregoes clarity and structure in presentation of results, for the sake of ``troubl[ing] received wisdom''\cite{fox2015inside} and towards generating new kinds of insights in research settings.

In this particular instance of DM, we acknowledge this aggregative, reductionist nature of TA, and attempt to reintroduce complexity by emphasising divergent events within common themes. We begin by coding the transcribed participant interviews, which involves the iterative labelling, grouping, and refining to collate a list of emergent themes. At this point we depart from traditional methods; rather than reporting on the themes as coherent or conclusive results as supported by corresponding quotes, we identify points of difference with respect to this theme, we present data (e.g. participant quotes, relevant theory) that appear on the surface to be in contradiction, and we attempt to reconcile such discrepancies by tracing them back to the greater context of the research assemblage. One unique feature of DM is the readiness to sit with such discrepancies. That is, to not discard them simply because they don't 'fit' within a greater organising structure. In our particular application, DM is characterised by allowing the data to evade necessary categorisation that would otherwise be imposed by traditional methodologies.

\section{Evaluation}

\subsection{Personalising Stable Diffusion}
Diffusion models provide a unique approach to generative modeling that marks a significant leap from its predecessors, namely, Generative Adversarial Networks (GANs) \cite{goodfellow2020generative}. Diffusion models employ a stochastic process that transforms a simple initial distribution into a complex data distribution. The core innovation of recent prompt-based image generators was the fusion of two independently powerful models -- CLIP \cite{radford2021learning} for text-based understanding and Diffusion Models for image synthesis -- into a cohesive system capable of generating images from textual prompts. 

For the purpose of this study, we opted to fine-tune the Stable Diffusion v1.4  model \cite{rombach2021highresolution} on a dataset of participant's artwork. This choice was driven by our overall intentions for this study; to investigate how TTI models can integrate into an artist's existing creative practice. Customisation of generative models is  supported in the literature as a proposed method of enhancing the co-creative process \cite{Ko2023}. 

\begin{figure*}
    \centering
    \includegraphics[width=1\textwidth]{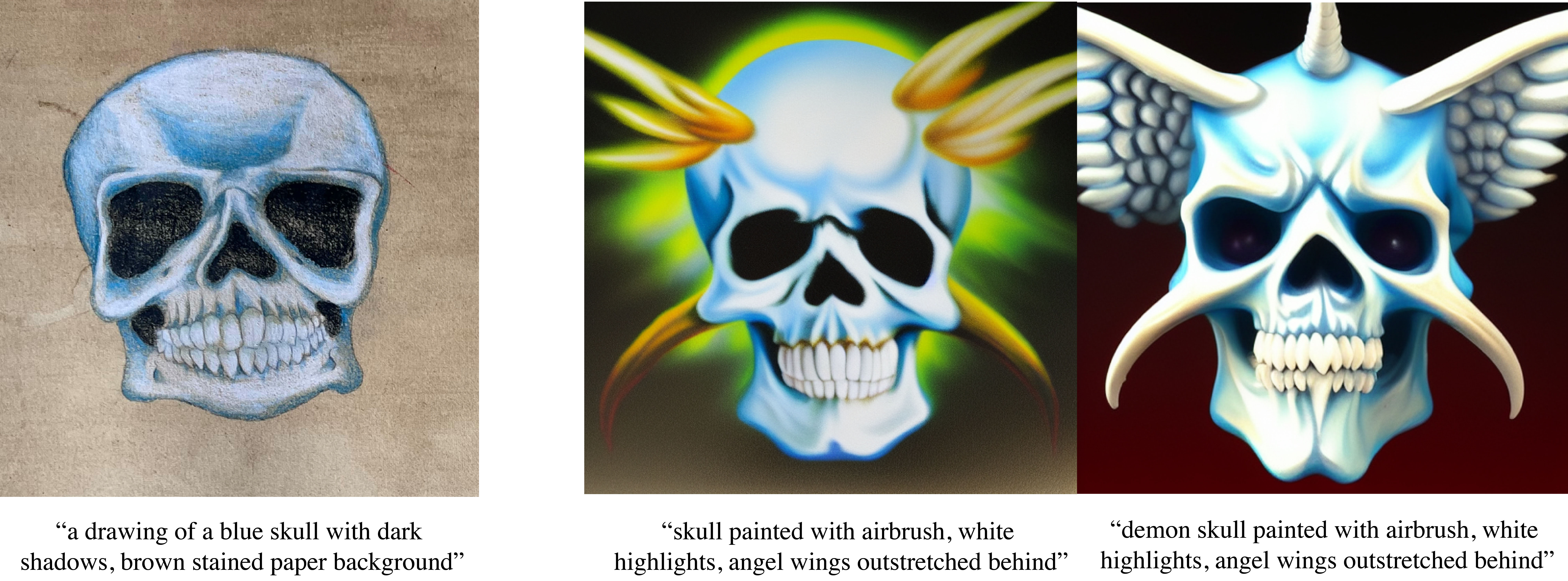}
    \caption{An image used in finetuning, with human-labelled caption below (left), beside two generated images from the same model with prompts (right)}
    \label{fig:zakskulls}
\end{figure*}

We finetuned the system using EveryDream \cite{everydream}, an open-source general purpose trainer that supports supervised learning. Participants were requested to submit 30+ images of their work. Each image was individually labeled with a descriptive caption (an example can be seen in Figure \label{fig:zakskulls}. The specific language choice was reached in collaboration with participants to ensure the correct terminology was used with respect to their creative domain. Finetuning was a relatively GPU intensive process, requiring 24GB of memory. The finetuned model was then made available to participants via a web interface \footnote{For the interface, we utilised the most widely used interface for Stable Diffusion  \cite{AUTOMATIC1111_Stable_Diffusion_Web_2022},}, along with the full set of captioned training data. Participants were instructed to look over the captions to familiarise themselves with the terms, prompt structure, and prompt ordering. Participants were also provided with basic instruction on using the interface, including a short guide to prompt engineering. 

Personalising our TTI model for each individual participant introduced an element of variability; the quality of the finetuned model depends significantly on the quality of the provided dataset, which differs considerably across participants. For this reason, we do not set out to evaluate the success of the finetuning method, rather to track the sustained engagement; cataloguing the ways in which participants naturally use the system without direct instruction. 

\subsection{Study Design}
To investigate the potential opportunities and challenges afforded by TTI systems, seven artists (4 male, 3 female) were recruited to collaborate with a personalised diffusion model over a two week period. Participants' ages ranged from 27 -- 35 with a median of 31. The primary selection criteria for our study group was artists who work with physical materials in their practice. This was in an effort to gain a different perspective from digital artists and illustrators who are often featured more prominently in existing studies on generative AI. Furthermore, we selected artists across a range of experience level, from those that engage in practice as a hobbyist, to those working professionally in their given specialisation.
The seven artists recruited work with visual media, including tattooing, design, illustration and painting.\footnote{The over-indexing of tattoo artists in our study is group is in part due to our particular selection criteria. Tattooing is a trade that is pursued by some visual artists as a way, in part, to achieve financial sustainability through their practice. This point is further elaborated on in Section \ref{sss:increased_efficiency}}

\begin{figure*}
    \centering
    \includegraphics[width=0.8\textwidth]{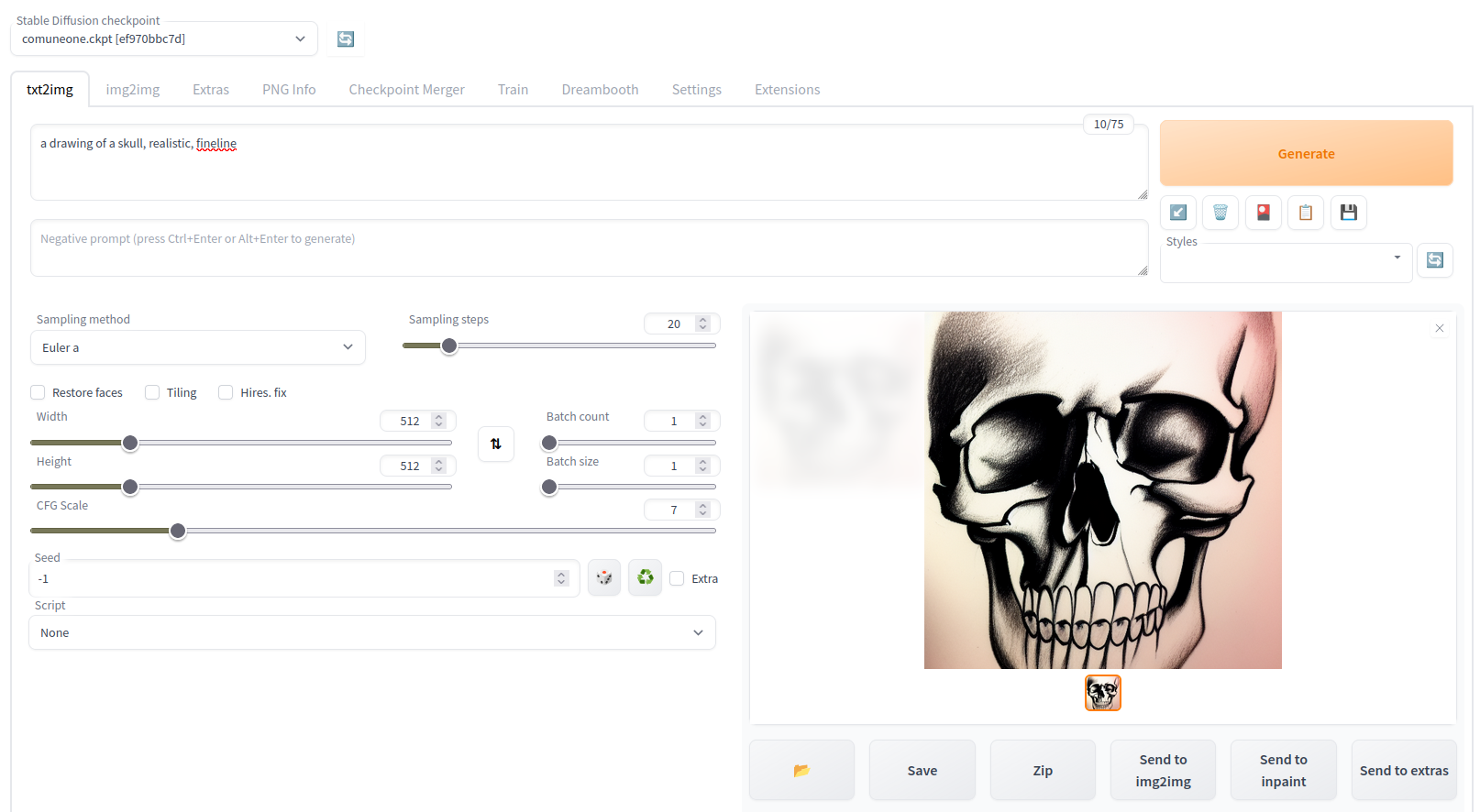}
    \caption{A screenshot of the web-app interface used for the personalised diffusion model}
    \label{fig:interface}
\end{figure*}

All participants have minimal experience in using prompt-based image generators. A brief background for each participant in this study is shown in Figure \ref{table:p}\footnote{Participants gave their consent for their names and the images they created for use in the study to be reproduced here. The study was carried out with formal approval from our University ethics committee.}. For the purpose of this preliminary investigation, the small study size reflects our chosen methodological approach; allowing us to probe deeply into variances and idiosyncrasies across and within individual participants, rather than striving to reach conclusive generalizations based on statistical prevalence.

Participants were given access to their personalised model via a web-app (Figure \ref{fig:interface}), and were instructed to engage via the standard prompt interface. Participants were asked to incorporate the personalised model into their creative practice and workflow with the freedom to integrate the AI model in any way they chose. There was no specified minimum or maximum use requested of participants. In addition, participants also had access to the generic Stable Diffusion base model. After the two week period, participants took part in a semi-structured interview regarding their overall experience working with the AI collaborator. At the conclusion of the study, individual responses were coded to identify common themes across participant experiences. According to the diffractive methodology as outlined in \ref{sec:diffractive}, the interview data was again read `through' these emergent themes, along with relevant theory, and contextualising information. The following diffractive analysis takes into account the differences between participant experiences, tracing the ways in which particular contexts and configurations might give rise to such differences; participants' unique viewpoint with respect to their artistic career, their current circumstances, and their attitudes towards technological adoption are taken into consideration. A spotlight on divergent participant experiences, opinions, and feelings provides a more measured and comprehensive account of the potential impact of generative AI on human creativity. We furthermore draw out a number of design considerations towards the future development of generative interfaces.

\begin{table*}
\begin{center}
\begin{tabular}{ | m{5em} | m{5cm}| m{2.5cm} | m{2.5cm} |} 
  \hline
  \textbf{Participant} & \textbf{Practice} & \textbf{Level} & \textbf{Experience}\\
  \hline
  Alex & Apparel Designer, Graphic Artist & Non-Professional & 3 Years\\ 
  \hline
  Andre & Illustrator, Tattoo artist & Professional & 3 Years\\ 
  \hline
   Christina & Tattoo Artist & Professional & 3 Years\\ 
  \hline
  Darien & Apprentice Tattoo Artist & Non-Professional & 1 Year\\ 
  \hline
  Georgia & Visual artist, Educator & Non-Professional & 10 Years\\ 
  \hline
  Kyle &  Fine artist, Tattoo artist & Professional & 15 Years\\ 
  \hline
  Zak & Painter, Tattoo Artist & Professional & 14 Years\\ 
  \hline
\end{tabular}
\caption{Background information of study participants, including self-identified title and experience level.}
\label{table:p}
\end{center}
\end{table*}

\subsection{Diffractive Analysis}
\label{sec:diffractive}

In the following diffractive analysis, we identify and unpack the shared and the distinct experiences across each participant.

\subsubsection{Creative Agency}
\label{sss:agency}

All participants in the study expressed a preference for their personalised model over the generic Stable Diffusion 1.4v model. In this particular study, personalisation of the model to a participant's cultivated style is intended to place the creative collaboration in the context of their existing artistic practice. Participants (Alex, Georgia, Andre) ascribed the model with the capability of providing an alternative interpretation of their visual aesthetic, and so they naturally assess the model output within this frame of reference. 

\begin{quote}
    Alex: ``It's like this very, like sensory sensory experience where you really feel like it's something you recognise, but then it's mixed with something you don't recognise and the experience is really beautiful''
\end{quote}

A number of participants described their enjoyment of the unexpected quality of the generated images. As discussed in Section \ref{sec:aicreativity}, the ascribed `creativity' of AI models emerges precisely from this handing over of control to the machine; the models themselves appear to have a creative agency in the sense that they offer imagery beyond what is directly asked of them. 

The discourse around machine agency has traditionally placed AI on a spectrum between a mere tool to independent agent \cite{epstein2020gets}. On one hand, having the AI model generate entirely irrelevant output is not conducive to a collaborative environment. Conversely, if the AI model merely replicates the training data, yielding imagery that appears overly familiar or repetitive, it, too, renders the collaboration fruitless. In the space of human-AI collaboration, there is a need to strike a balance; to have the AI take clear direction, and at the same time to offer something novel.

Identifying the optimal balance of agencies was a challenge for many participants. This was discussed by way of the CFG parameter \footnote{Classifier Free Guidance (CFG) determines how closely the model adheres to a given prompt, and is represented by a score that estimates trade-off between output quality and diversity \cite{ho2022classifier}} that controls how much a given prompt has influence over an output image.

\begin{quote}
    Georgia: ``I remember reading the scale, and it was always the ones that I liked the best were towards following my prompt religiously, but allowing it a little bit of leeway, like it was never fully up the scale. Otherwise I think it became a bit too basic, I suppose. And it didn't have those interesting qualities that the AI would like bring to the table''
\end{quote}

\begin{figure*}
    \centering
    \includegraphics[width=1\textwidth]{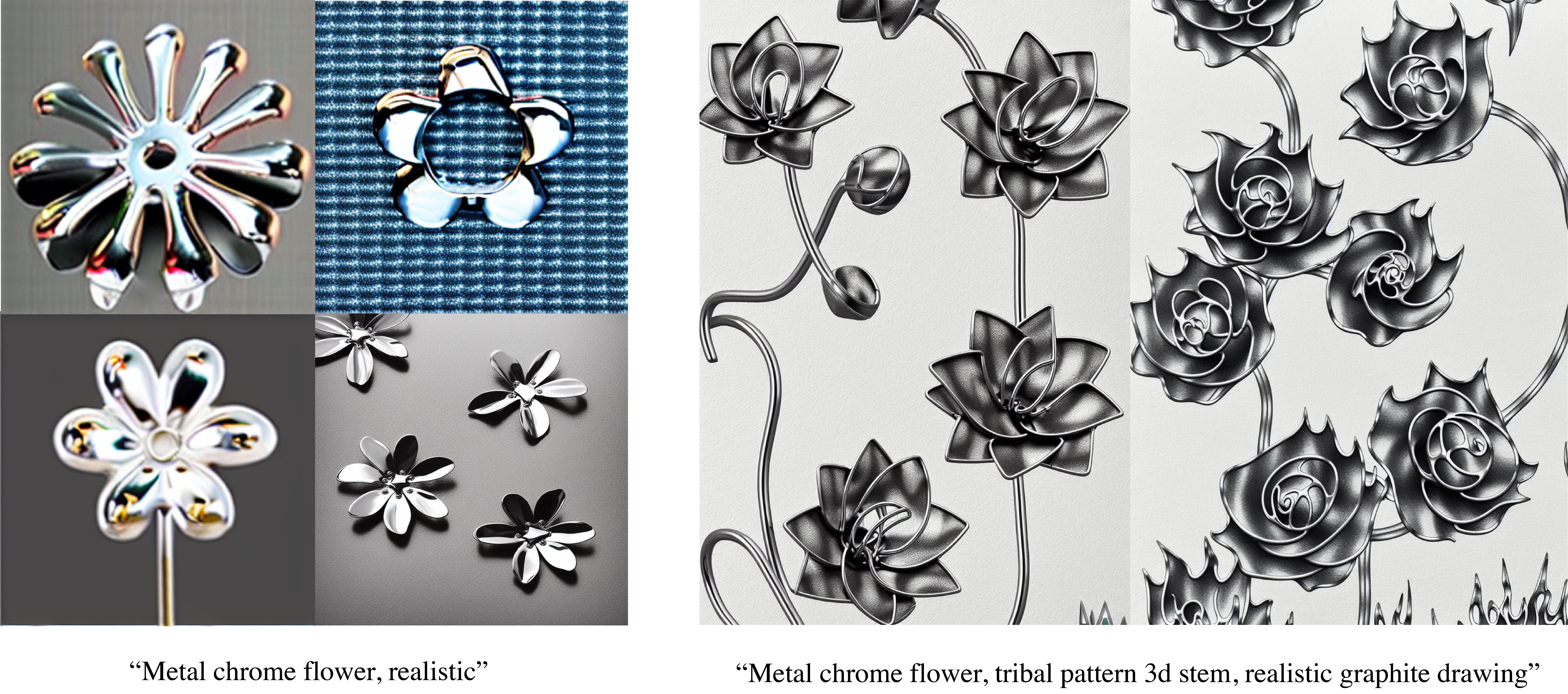}
    \caption{A selection of early attempts to generate a metallic flower (left), final attempts to generate a metallic flower (right)}
    \label{fig:metalflower}
\end{figure*}

Georgia touched upon the value of AI models in their ability to offer a unique artistic perspective; to be treated entirely as a tool is to undermine their creative agency, and to compromise the model's ability to offer creative inspiration:

\begin{quote}
    Georgia:``I do want the AI to take a little bit of the lead because it is a bit more interesting. Otherwise, it's like, what's the point? I'll just do the work myself.''
\end{quote}

On the other hand, Kyle expressed a preference for the AI to generate exactly what they had envisioned in their head

\begin{quote}
    Kyle: ``it was also just because I maybe had a preconceived idea on what I was looking for the machines to produce almost, you know what I mean? So it's like if I'm putting in, you know, a metal flower and it's not metal, it's a bit like, this isn't quite the idea generation I'm after [...] I just wasn't getting anything that was sort of resembling what I was looking for.''
\end{quote}

Kyle generated 20 `metal flower' images that they were not satisfied with. After increasing the image, resolution and honing in on the prompt, Kyle was finally able to reach output closer to their personal style (Figure \ref{fig:metalflower}). Nevertheless, Kyle describes this process as tedious and unsatisfying (\textit{``I think it's definitely not quite as engaging as it would be if I'm just trying to sketch ideas out on an iPad or something like that, you know''}), with resulting images falling short of their standard of quality (\textit{``They're still not quite there for me'}).

Both participants compared working with the system to their existing workflow, developed as the basis of their creative practice. Interestingly, they come to the same question from opposite sides: If the AI system is simply mimicking ones work, then what value does engaging with the model add to ones practice? In contrast, if the AI system doesn't generate precisely what the participant is imagining, then, again, it is perhaps easier not use the model at all. 

We highlight the discordance between these two participants experiences in order to stress the following point: There is no ideal balance of agencies between human and AI, rather this is entirely context dependent. The perception of machine agency (as too dominating, or as too muted) depends upon the particular needs of the human participant. Georgia expressed interacting with the AI model as a way to gather novel visual ideas; ``\textit{ I think I would use it early on in the process of making a work, or if I was sketching around [...] definitely a tool for early artistic process and experimenting}''. Whereas Kyle's primary interest in the model was to see if it could produce convincing variants of his work; \textit{``the thought of it maybe being able to sort of recreate my own work, I guess is definitely something that I find pretty interesting''}. The source of this difference in usecase can be attributed to each participants relative professional context. Kyle is a working artist who expressed an interest in whether automation may increase his productive, and hence economic, output. Georgia, on the other hand, presented with no direct economic incentive, rather an interest in how AI could influence or inspire their practice.

Georgia approached the system as an idea generator; a sort of mirror reflecting back their personal aesthetic in a novel way, with the goal of experimenting and pushing their style. For Kyle, on the other hand, one expectation of the model was to reproduce their work convincingly. As such, its performance was measured against an entirely different set of standards. The distinction between AI as an ideation tool, and AI as production tool, reemerges several times throughout the following analysis. With respect to the above discussion, however, we see that AI for ideation necessitates contribution of machine agency, whereas AI for content production requires a greater degree of human control. 

\subsubsection{Ideation}
\label{sss:ideation}

A majority of participants enjoyed engaging with the model for the purposes of ideation. This stage generally took place at the beginning of their normal workflow, as a catalyst to creative thinking. As discussed above, Georgia primarily used the model to generate ideas before redrawing elements (Figure \ref{fig:georgiapainting}) of the model output: \textit{``If I just save these on my computer, I'll probably never look at them again. But if I recreate them by hand, they'll be able to stick in my brain a little more''}.

For Alex, engaging with the model was done primarily in the stages of design conceptualisation. In this case, the value of the model was less to do with whether the output was perceived to be 'good' or 'bad', but rather as a tool to encourage divergence of ones visual style :

\begin{quote}
    Alex: ``I feel like sometimes you need to kill your baby to like, you know, get to other places. And like, putting into AI is kind of like a really big version of just killing a baby. It's just like, take my baby and just spit out heaps of stuff. And maybe I'll like it, or maybe I'll hate it, and there's some things that it makes that I hate. And like, that's cool. I make things I hate as well.''
\end{quote}

Divergent thinking has been explored extensively in the literature on human creativity and creative practice \cite{runco1991divergent,runco2014neuroscience}. Neuroscientific studies corroborate the role of divergent thinking as a mental exercise that can stimulate creative outcomes, thereby validating the value attributed to the model as a tool for encouraging this kind of exploratory process. In this light, we see that the model itself doesn't necessarily require greater degree of 'controlability' or accuracy in order to facilitate ideation. The capacity for generative AI systems to foster creative inspiration lies precisely in their perceived unpredictability. The inherent qualities of such systems lend themselves naturally towards divergent perspectives, simply for the reason that they exert a new kind of creative agency, and in doing so generate new creative possibilities.

Viewing diffusion models as an artistic medium furthermore warrants a highlighting of their unique characteristics, rather than attempting to conceal them, or to imitate traditional mediums (i.e. to generate a convincingly human painting). In light of the above discussion on machine agency; ``it is precisely the resistance of the medium to being \emph{moulded} that leads to it's true creative potential'' \cite{mccormack2023writing}. In this case of Stable Diffusion, this unique quality includes errors in the systems ability to produce legible text, or correct anatomical structure (see Figure \ref{fig:errors}).

In contrast, other participants articulated limitations regarding the model as an ideation tool. The primary critique revolved around the machine's inability to independently generate conceptual ideas, suggesting that the creative onus is left for the human participant to determine:

\begin{quote}
     Andre: ``I suppose I had this idea that maybe it would make life a whole lot easier, and it would kind of generate ideas for me, but, you know, you're still the one that has to input the ideas. So there is still the human element, which is, arguably, the most difficult to create, you're still kind of like left on your own to do it''
\end{quote}

Here, Andre provides a critique of the prompt interface, rather than of the aesthetic qualities of the model itself. The design of such an interface presumes a collaboration in which the human participant holds a preexisting mental image or vision, and that the model's role is to realise that vision. Yet, idea generation occurs largely before this stage of the creative process. In reference to the two modes identified in \ref{sss:agency}, ideation and production, it is clear that the current interface lends itself primarily towards the latter. TTI systems do not provide conceptual suggestions, only responsive imagery. Andre's comment reflects a broader feature of mainstream TTI systems and LLMs alike; they are designed precisely to be \emph{general}. The human participant, then, is left to determine the application of the technology. While being an intentional and pertinent feature of generative AI systems, it does not bode well for the purpose of ideation.

A number of participants expressed frustration with the prompt-based interface, and instead began to enter obtuse or poetic prompts with no singular expectation of the output; opting to engage in a creative dialogue with the machine. Darien, for instance, described enjoying the interaction more when using non-literal text prompts:

\begin{figure*}
    \centering
    \includegraphics[width=0.9\textwidth]{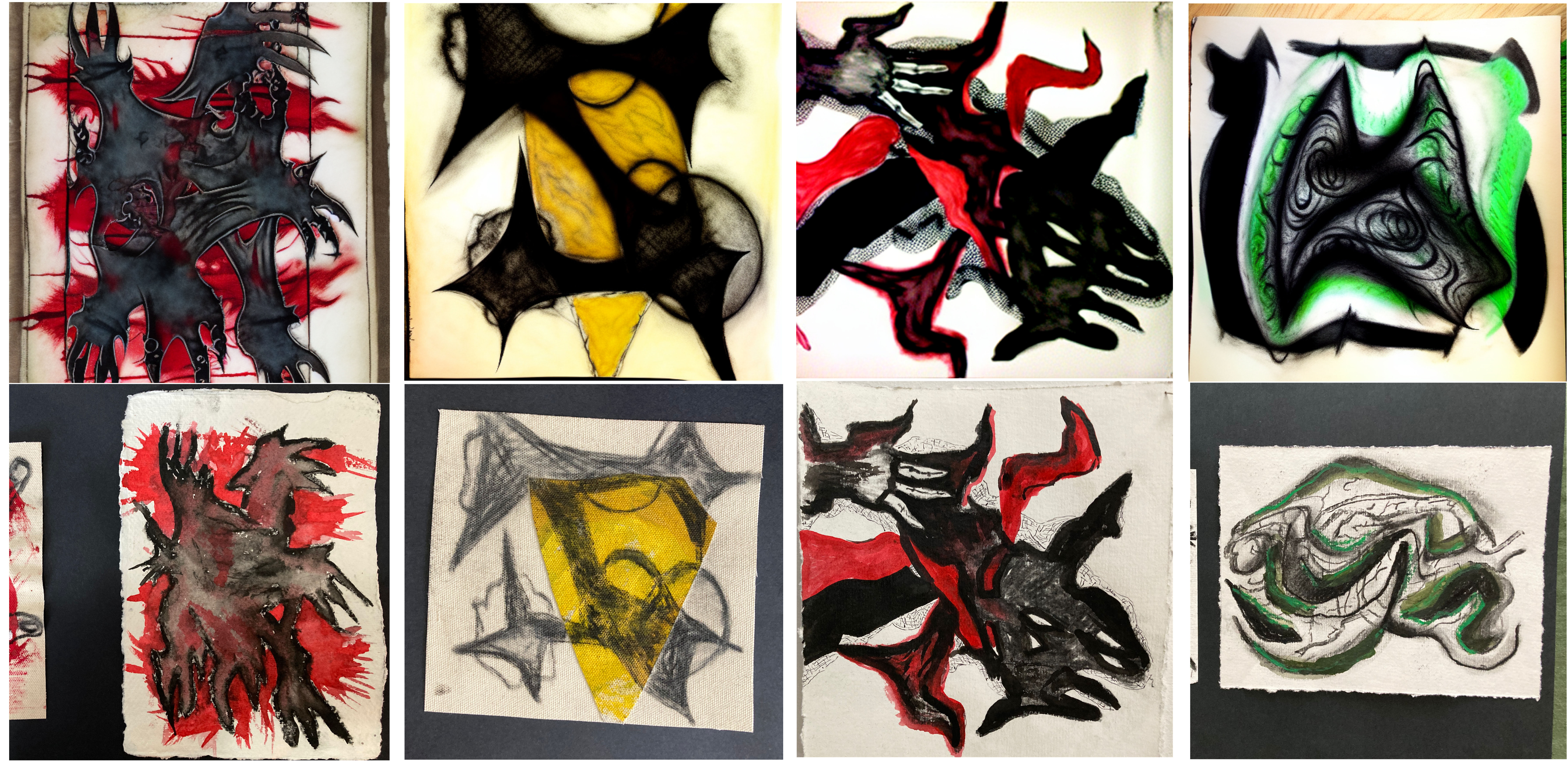}
    \caption{A selection of AI generated images (top) recreated by hand (bottom) as a part Georgia's creative experimentation.}
    \label{fig:georgiapainting}
\end{figure*}

\begin{quote}
    Darien: ``When the prompt itself was quite abstract, then it came up with things that I actually found visually interesting... Because when I draw things, a lot of the time, I'm not so much necessarily giving myself the instruction of like, okay, I'm gonna draw, you know, a flash sheet with a skull. You know, it's more like, I'm in a certain mood, or like, I feel like celebrating something. Yeah. Like a feeling or like a song or, it's just like, some kind of abstract feeling related to something else. And then I'll draw something that I think represents that thing.''
\end{quote}

\begin{figure*}
    \centering
    \includegraphics[width=\textwidth]{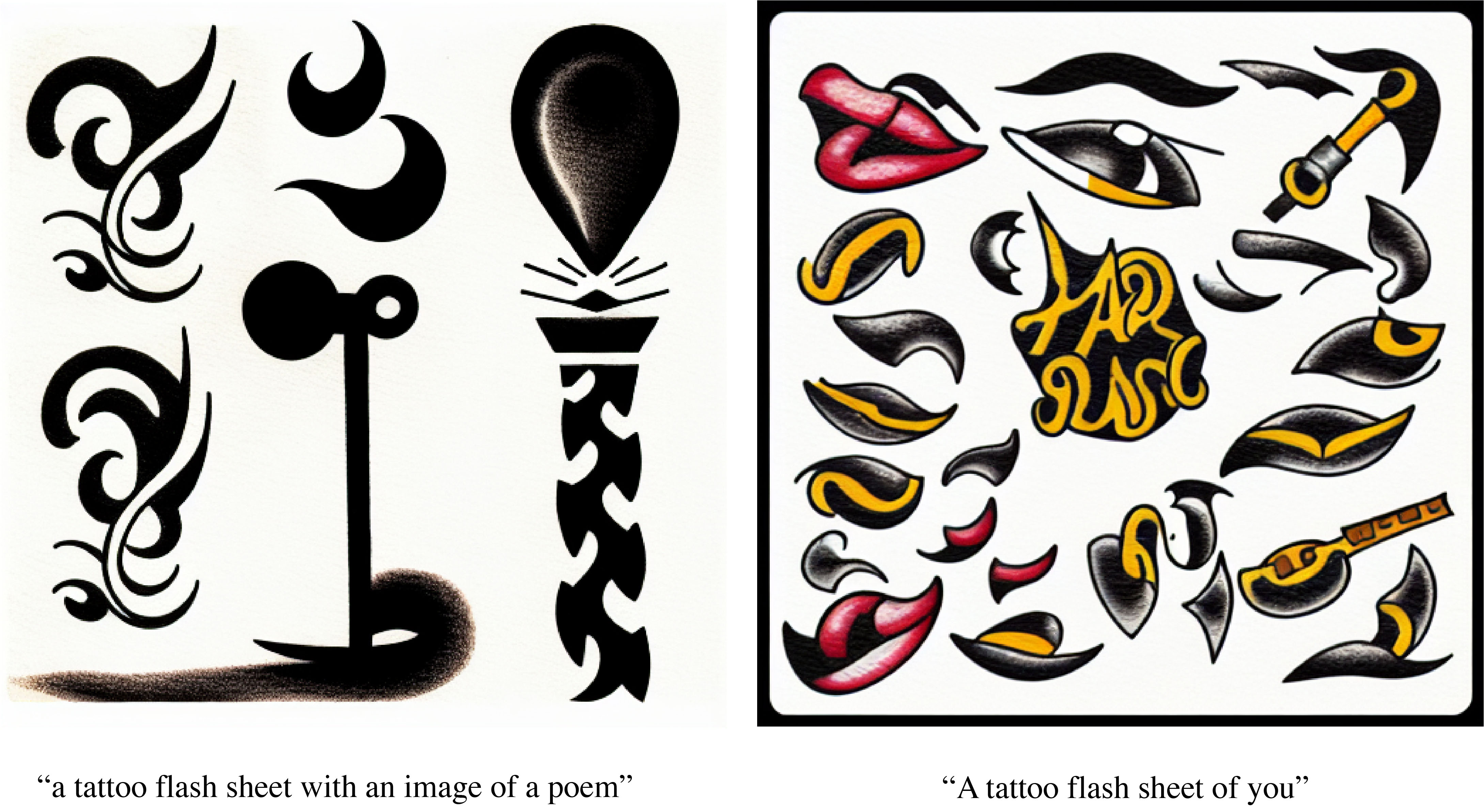}
    \caption{Generative images with prompts displayed underneath; an example of non-literal prompting}
    \label{fig:nonliteral}
\end{figure*}

The generated image produced with a non-literal prompt stood out to Darien as one of the few images they liked;  ``I was kinda into this one'' in reference to Figure \ref{fig:nonliteral} (left). 

Synthesising insight from these differing experiences, we gather that generative systems certainly do encourage divergent thinking due to the quality of the medium. Nevertheless, the specific interface design of TTI systems means that human input is necessary. Moreover, it is not just human input, but rather human guidance from beginning to end. For the purposes of ideation, perhaps the system could benefit from a greater degree of agency; a creative system that is able to offer divergence unprompted. 

\subsubsection{Plagiarism}
\label{sss:plagiarism}
 
When examining models such as Stable Diffusion, the ethical implications of data sourcing cannot be overlooked. Most participants expressed some awareness of the ongoing ethical debate surrounding the dataset scraping and subsequent training of the Stable Diffusion. While these models present groundbreaking advancements in technology, they source their training data through web scraping methods that bypass artists' consent. This practice has been the subject of intense debate, questioning not just the legality but the ethical validity of using creative works without permission. The discussion extends beyond copyright infringement, touching upon moral rights, the decontextualization of art, and the potential exploitation of unrecognized creators \cite{data_laundry,mirowski:2023}.

Artists in particular have become increasingly vocal regarding the perceived threat to their livelihood and craft posed by such technologies \cite{jiang2023ai}. While echoing a similar sentiment, a number of participants identified parallels between the plagiarism underpinning TTI models and the culture of `referencing' in particular creative industries. Georgia, for instance, expressed a feeling that their participation was ethically dubious: 

\begin{quote}
    Georgia: ``when I was younger, I used to print out, like drawings and then trace over them. You know what I mean? And I guess that's, like, it's sort of replicated the same feeling like, Oh, my God, I'm just like, copying something that's not really mine. And it feels a bit gross, I suppose, a bit plagiaristic.''
\end{quote}

Drawing inspiration from and referencing other artists' work is a standard practice across all creative industries. Referencing, citing, and pulling inspiration from other sources are all integral to developing and carving out ones unique personal style. Kyle draws parallels between this aspect of creative work and an AI model trained on human art, building something of an argument in defense of generative models.

\begin{quote}
    Kyle: ``when we create art, sort of, organically or whatever, we're still using a whole tonne of reference that we've already got, whether it's inside our head, or we're like looking at another picture. I guess in that sense, it's still quite a similar process.''
\end{quote}

Lingering on Kyle's phrasing -- \emph{"organically or whatever''} -- we see discourse unfold around the fluid concept of what is and is not deemed `natural' in the face of technological progress. In Kyle's practice, image aggregators such as Google Images and Pinterest are utilised on a near daily basis to source inspiration and reference imagery; a repertoire of tools that have fundamentally transformed Kyle's trade of tattoo artistry. There is no line to be drawn here between organic or non-organic, for the line is always moving. What is considered `organic' in today's lexicon would have been alien half a century ago. What may be classified as 'artificial' now is likely to be subsumed into future definitions of the human.

Kyle conflation of an AI system's training dataset, and a lifetime of human experiences, of course, falls short in a number of areas. As Darien argues, one key difference lies in the inability to properly acknowledge and cite human work when pulling inspiration from AI generated imagery. 

\begin{quote}
    Darien: `` I've always respected the idea of sourcing imagery, like going into the person directly, an AI just seems like so impersonal, and so removed from that idea, that doesn't sit right.''
\end{quote}

As Christina similarly notes, \emph{``there's a bit of history behind every image''}. Without the ability to acknowledge the lineage of visual and aesthetic development, Christina argues, \emph{``you can't make something new''}.
Nevertheless, to flag TTI models as plagiaristic implicitly assumes that traditional practices employed in the creation of artistic works are ethically sound. In some cases, these practices abide by a legal framework such as copyright laws. In other cases, there has emerged a social norm regarding what does and doesn't count as plagiarism, which is commonly enforced among social groups through often unwritten and fluid rules.

\begin{quote}
    Darien: ``it's something that is worked out personally. Like, so many people are using the same symbols and things, but you can just sort of tell when someone's put certain things in combination, and it's trying to emanate the same thing in the same way as someone else. And can just look at it and be like, are you copying?''
\end{quote}

Meanwhile, in the fashion industry, Alex highlights the pervasive practice of large brands pilfering designs from smaller creators.

\begin{quote}
    Alex: ``some brands, just copy people on the internet, and then make it, and that's their workflow. You know, I don't agree with that. But some corporations just take an interns work, and then just make it their way, And then they're like, yeah, it's our workflow''
\end{quote}

In a similar vein, Zak notes that contemporary art practices also see an outsourcing of the material production to often uncredited and poorly compensated workers:

\begin{quote}
    Zak: ``there's so many contemporary artists out there that just give prompts to the artist assistants, and they make things and that's like, you see so many shows and it's like that''
\end{quote}

In a recent and notable legal dispute, the sculptor who fabricated the artworks attributed to Maurizio Cattelan filed a lawsuit, alleging misappropriation of credit for the creations\footnote{https://www.artnews.com/art-news/news/maurizio-cattelan-authorship-lawsuit-dismissed-1234633671/}. Despite the legal challenge, the case was ultimately dismissed, highlighting the complexities of authorship and creative ownership within the art world.
Alex suggests that with plagiarism already rife in a number of creative industries, the introduction of TTI models are in fact democratising creative production:

\begin{quote}
    Alex: ``if you're going against people who have more money than you and more time than you, or like more people than you, you could use that as a tool to really even the playing field a little bit''
\end{quote}

The difference in perspectives across participants perhaps shines light on the preexisting societal context that these technological advancements are affecting. Established names may profit from the work of lesser known artists and interns, often without credit. As alluded to by Alex, TTI models allow for the converse. Emerging artists are now granted a new kind of outsourcing power, which has potential to shrink the gap caused by an artists financial means. In the current debate around plagiarism of the dataset\footnote{https://www.newyorker.com/culture/infinite-scroll/is-ai-art-stealing-from-artists}, we see a disproportionate representation of voices whose work is public enough to have been scraped, and who consequently have a financial stake in the matter. The participants in this study comprise of young emerging and professional artists who perhaps have little to lose from the development of such technology, and proportionally more to gain.

\begin{quote}
    Alex: ``The art scene is very notorious, like burning yourself out, dedicating yourself, and I have done that, that's happened to me. I've just dedicated all my time and energy into it. And it's, it's because I'm passionate about it, like I'm okay with that. But they just really, really, really grind it out of you. So I think AI has the, I don't know, has that sort of potential to ease that strain a little bit''
\end{quote}

\subsubsection{Efficiency}
\label{sss:increased_efficiency}

When asked to envision how they would ideally utilise an AI image model in their practice, a number of participants (Christina, Andre, Zak, Kyle) describe simple tasks that they would happily have outsourced, purely for the sake of saving time. Andre notes that creative professionals don't exist outside of the bounds of profit-driven society. 

\begin{quote}
    Andre: ``We're no better than the people in in like, high corporations that use [generative AI] to kind of optimise so they can fire people because the computer can do the work that they can do. You know, we're no different to that. We want it so that we can optimise our time. So we can go and sit back in the staff room and chat''
\end{quote}

Kyle outlines a number of technical interface-level improvements that would make a tool such as Stable diffusion worthwhile; if it were integrated into their existing suite of creative tools, if the model could generate high-quality images instantly, and if it could respond collaboratively to work-in-progress sketches. Christina furthermore lists highly domain-specific features that would save time in their particular practice, such as changing line width of a drawing, or creating a stencil of a drawing with one click.

These suggestions are often fueled by a desire to maximise time spent on more fulfilling work. As Andre expresses, \emph{``maybe I can get faster at the the most inane part of the job''}. Or alternatively, as Zak notes, to simply make time to produce \emph{more}.

\begin{quote}
    Zak: ``I can spend that time to maximise profit, make more designs. I mean, the ideal answer would be leisure. But we don't exist in a world that allows that, to that extent, yet''
\end{quote}

Notably, the participants who were more inclined to adopt time-saving technologies, were those who have managed to transform their creative practice into a full-time trade, all of which are tattoo artists.  This is contrasted with the remaining participants (Georgia, Alex, Darien) who have yet to achieve financial sustainability through their practice. This difference speaks to the difficulty of making a living from ones art, with artists often supplementing or centering their practice around trade. With the tattooing industry currently booming \footnote{According to IBIS, the tattoo industry in the U.S. grew 8.1\% a year on average between 2018 and 2023 \cite{ibistattoo}}, it presents as a desirable and marketable skill for artists. This serves to remind us of the ways in which creativity must be co-opted towards the production of value in order to sustain itself.

\begin{quote}
    Kyle: ``For a lot of my stuff, it goes hand in hand, most creativity ends up being somewhat monetized.''
\end{quote}

According to \cite{larsen2014compulsory}, creativity is in fact a major driving force of a capitalist economy. Within this framework, creativity is ceaselessly appropriated to generate economic value; it becomes ``difficult to know precisely where the individual use value of creativity stops and the exchange value of the original and creative talents begins'' \cite{larsen2014compulsory}.
Likewise, the advancement of generative AI models, particularly those that automate creative tasks, is propelled by capitalism’s fixation on economic growth. The intention for these systems appear to be not so much to enrich our lives in a creative or meaningful manner, but rather to perpetuate this cycle of accelerated production and consumption driven by capital.
In attempts to 'automate creativity', the recent advancement of generative AI systems echo the economic revolutions of the 19th and 20th centuries; in which manual labour was outsourced to the machine \cite{agbaji2023perceptions}.

Returning to our categorisation of AI towards ideation, and production, respectively, it is not so clear where one draws the line between creativity as subjective experience, and creativity as a productive force. The two modes are inextricably linked; a tool for generating creative innovation is ultimately used towards the production of value. Moreover, maintaining a financially viable practice enables artists to further refine and expand their creative work.

\subsubsection{Identity and Authorship}
\label{sss:authorship}

As previously mentioned, all participants expressed a fondness for their personalised version of \emph{Stable Diffusion}, as opposed to the generic models. They describe the unique interactions it allowed and the novel perspectives it offered into their own aesthetic and creative identity. The customisation, as such, offered a mirror to their artistic identities, prompting introspective thoughts and fresh viewpoints on their own style.

\begin{quote}
    Alex: ``There is this kind of essence of like, when I see an AI image, I know it's not mine. I know, I didn't create it like, technically, even though it has my style on it, there's this little distance between it''
\end{quote}

The personalised model was perceived by some participants to capture a small piece of their artistic identity; extending beyond their 'natural' capacities and into a kind of immutable, digital artefact. As expected, this extending of the self into AI provoked both delight and concern across participants. More often than not, simultaneously.
Christina was initially impressed with the capability of the model to produce convincing work in their style. Yet, after a few days of use, the participant describes an unsettling experience in which they felt as if the system was improving the more they engaged with it.

\begin{quote}
    Christina: ``It felt like I was feeding it''
\end{quote}

This led Christina to stop using the system for the remainder of the two week period. These fears were, in part, driven by a lack of understanding into the internal working of such TTI systems. In reality, the system appeared to be improving simply because Christina was becoming more adept at the language and art of prompting -- no further training on the participant's creative work was used beyond the initial sample.

Christina's withdrawal was symptomatic of a broader anxiety emerging concerning the sanctity of a creators role in the age of generative AI \cite{hutson2023poetry}. This concern is accentuated in a landscape where digital dissemination of artwork is not only common, but increasingly necessary for career success. The accessibility of tools like Stable Diffusion present the risk of anyone appropriating and commodifying an artists individual style, undermining its originality and, by extension, its monetary value.

All participants articulated reservations in claiming authorship of the images generated by the model. A number of participants ascribe this to the lack of effort and physical labour invested into the production of the final image.
\begin{quote}
    Kyle: ``it hasn't come from my hand [...] I'm still not the one who actually created it''
\end{quote}
\begin{quote}
    Zak: ``I wouldn't probably ever use a design fully from there, even if it could render my style faithfully, sort of like wrap it, put a bow on it, and it's done. I still would have to redraw it, put some of my own hand in it to feel like it was mine.''
\end{quote}
\begin{quote}
    Georgia: ``It felt like, it wasn't mine, because I didn't conceptually come up with it. I didn't experiment with the techniques. I didn't experiment with the mediums. I didn't sketch it out. It just kind of popped up on my screen.''
\end{quote}

For these participants, claiming authorship would require them incorporating the output into their creative practice \footnote{A recent inquiry by the U.S. Copyright Office into Generative AI ruled that a piece of art created by AI is not open to copyright protection \cite{ai_copyright} without significant modification by a human}. For example, redrawing the image by hand, using only elements of the generated output, or generating variations of an original drawing, were just a few methods envisioned by participants for collaboration. For all participants, the generated image did not qualify as an artwork in its own right. Rather, engaging with the system over time presented potential for creative collaboration.
For artists with an established practice and workflow, generative AI and TTI systems could be envisioned as an extra step in their process, rather than replacing a component of it. 

\begin{quote}
    Christina: ``The way that I'm using it, as opposed to like, it being perfect. I'm almost like, I'm cool with it not being perfect. Because I wouldn't want to rely on it. I could definitely figure out a place that it fits into the process''
\end{quote}

This sentiment resonates with the extensive literature promoting co-creative AI as the favoured paradigm for designing creative systems \cite{rezwana2022designing, d2015heroic}. As discussed in Section \ref{sec:cocreative}, co-creative frameworks encourage the enhancement and extension of human creativity, rather than the outsourcing of creative production to the machine. In contrast, the stream of research into Computational Creativity \cite{colton2012computational} has explored precisely this possibility; whether a machine is or can ever be considered as author, creator, as artist in it's own right.

By virtue of our study design, participants were not prompted to conceptualise the machine as one or the other. Yet, each participant engaged with the system as a tool, a creative assistant, but also as an extension of their identity; a means to increase and enhance their creative output. Notably, the system was conceptualised as an extension of their creative capacities rather than an independent and intentioned collaborator, with imagery generated by the model to be ultimately subsumed into the participant's greater artistic intention. 

This raises an important point for co-creative systems; Human-AI creative collaborative frameworks must not be built solely from our understanding of human-human creative collaboration, for the machine possesses an entirely different ontological and epistemological sensibility. To understand the models intention behind the generated output was often irrelevant to the participants, or rather, it is impossible for them to comprehend without falling back on anthropomorphic illusions. In the development and analysis of co-creative systems, reverting to anthropocentric conceptualisations leads us to overlook the way in which these new generative technologies are shaping human creativity. These systems do not simply serve as replication of (human) creative agents, they extend upon human capacities for creativity \cite{hernandez2019ai}. Human and AI are not independent and analogous participants, but as mutually constitutive aspects of a whole. The human is in the machine (built for human purposes, trained on human data), and the machine is in the human (a technological extension of human capability, perception, and identity).

\subsubsection{Materiality}
\label{sss:materiality}

A recurring theme emerged around the tangible and material nature of the artists' practices. All of the artists who participated in this study engage in a creative practice that is founded in physical materials. Often times this material component is the most crucial; often associated with one's mastering of a 'skill' or 'craft'. This includes, drawing, tattooing, painting, sewing, sewing, and designing. In each case, participants describe an inability for the TTI model to comprehend the embodied knowledge that underpins creative production, alluding to an ontological difference between the embodied, materially-grounded knowledge in their creative practices and the abstract computational underpinnings of the AI model. In some cases, participants highlighted the impossibility of certain machine-generated images being made in the real world.

\begin{quote}
    Georgia: ``you can actually kind of tell it's AI because some of the images are almost impossible to create.''
\end{quote}

Christina similarly puts into question whether they, personally, would be able to recreate a generated image, even though it was originally trained on their hand-drawn work. 

\begin{quote}
    Christina: ``This is just straight visualisation. Like, you can see where it would be. It feels like it almost cuts out like the steps in between. But then it's like, can I still reproduce that? I don't know.''
\end{quote}

In one sense, the model is detached from the `material' world in that it appears unable to comprehend it. This comes as a detriment to its perceived creative potential, while at the same time minimising participant's fears around the colonisation of their creative pursuits.
When discussing perceived fear of automation and threats to their livelihood, Kyle expressed comfort in the knowledge that an AI model could not encroach on this space as their practice is rooted in working directly with physical materials; a territory that generative AI has not yet invaded.
    
\begin{quote}
    Kyle: ``It is a slightly weird feeling, I think. I'm possibly fortunate in the fact that the work I do is still quite hands on, manual, at least for now, until we've got, you know, robots tattooing people [...] I think it's still possibly a little while off. I don't think I'm necessarily against it''
\end{quote}
Again, we witness ambivalence within individual participants. In Section \ref{sss:authorship}, Christina experienced feelings of anxiety and paranoia (\textit{``it wigged me out'''}) after finding the system to be incredibly sophisticated and useful. On the other hand, Kyle's fears of automation were eased upon discovering that generative AI is not as advanced as they had expected. These apparent contradicting reports are, in fact, pointing towards the same thing; that AI can be neither all `good' or all 'bad'. Striving to develop systems that are `better' in one domain inevitably brings about potential negative implications in another. Instead of centering evaluation only on whether an AI system is `good', we must continue to probe precisely how, why, and most importantly, good for who?

Finally, for several participants (Andre, Zak, Christina) the value of art was deeply rooted in the \textit{knowing} that it was made by a human. In particular, the visual qualities that in a sense gave away that the work was done by hand; a shaky line, uneven application, or irregularities in form and design, become a window into the artists state of mind. 

\begin{quote}
    Zak: ``It's kind of like sign painting, you know, like, you see, you see a hand painted sign, I personally respond to that with a human, kind of, like a visceral human response, my eyes averted to it''
\end{quote}
\begin{quote}
    Andre: ``You kind of are looking for like the human errors, you know, you're kind of doing a scan. And that's when you actually realise how much skill has been involved in producing this piece''
\end{quote}
\begin{quote}
    Zak: ``our imagination is kind of centred within the digital realm, but we can still apply things badly. And that makes us human''
\end{quote}
    
And in conjunction, the value of creating art for our participants is founded in the process itself.

\begin{quote}
    Andre ``it's like this, the satisfaction that you get from like, putting in however many hours it takes to create something, the journey is the best part, the result are just the icing on top, you know, that you get to see something that represents the hours toiled, and the path that it took to actually get you to create that thing''
\end{quote}
\begin{quote}
    Georgia: ``It didn't give me the feeling of creating work that I usually get, like, it didn't give me the drive or the, I don't know what to call it, like the artistic expression, the spiritual expression, because it wasn't driven from me, it was driven by something else.''
\end{quote}

All participants expressed a similar sentiment around the lack of creative satisfaction granted from engaging with the prompt-based interface. This appears to have little to do with the quality of the output, but the nature of the interaction; one-shot queries, mediated by language, and confined to a screen (Darien: `\textit{`I'd just rather go and paint than sit at a computer''}). According to our participants, creative satisfaction emerges from the invested time and effort necessary to produce a novel or complex work, to master a skill (Christina: \textit{``mastership of craft''}), to express and articulate ones unique perspective of the world.
While integrating generative AI into an artist's workflow could very well automate tedious tasks, thereby, freeing up time to focus on creative work, this raises pivotal questions. Does automating creativity merely serve economic ends? And if so, what impact will this shift present towards the subjective experience of creativity?

The study led the participants to become more curious about AI technologies, their recent advancement, and in some cases, came to consider it as more of a threat. From the outset, all participants explain that they initially volunteered for this study because they had a curiosity about AI technologies, its capabilities, its potential as an assistive tool, perhaps even fueled by a morbid curiosity; to see if it was possible to automate some part of their artistic practice.
To this end, the model at its current state fell below the mark for all participants. But of course, with the rapid technological advancements seen in recent years, the feasibility of such is virtually guaranteed, and might even be realised by the time this research study is published. 

\section{Discussion}
\label{sec:discussion}

Our study identified five emergent themes for artists using TTI systems: creative agency, plagiarism, efficiency, identity \& authorship, and materiality. Through a diffractive analysis, we identified two dominant modes of use; \emph{AI for ideation}, and \emph{AI for production}. In this section we use the results of our analysis to synthesise a set of basic design considerations as well as ethical implications for future image-based generative AI systems.

\subsection{Design Considerations}
\label{ss:design_considerations}

We identified two modes of collaboration, ideation and production, that carry a different set of considerations towards future interface design.
Firstly, the two modes call for different levels of machine agency.
For the former, participants required more creative agency from the system to help them ``imagine'' ideas and visual forms. For the latter, more control was required for the output to match the participant's already imagined expectation. 

Secondly, we identified a number of limitations of current TTI systems towards both modes. In the case of ideation, we argue that the generality of the prompt interface hinders creative ideation. Moreover, TTI systems are limited by what can and can't be specified linguistically \cite{mccormack2023writing}; something that doesn't affect traditional visual media practices. More recent developments in TTI software have begun to recognise this limitation. MidJourney, for example, allows the use of images in addition to prompts as part of the generative process (e.g. remixing images, outpainting). Despite these enhancements, instructional language is still necessary as the mediator. In the case of production, limitations include the time taken to generate high-quality results, and issues around interface accessibility.

\subsubsection{AI for Ideation}
We summarise the design considerations necessary for using TTI systems for Ideation as follows:

\begin{itemize}
    \item {\em Exploit the unique properties of the medium:} Improving on the quality of the model's output does not necessarily benefit the creative potential of the system as a whole. Striving to precisely replicate reality exactly as it is, may not be the best form of inspiration. Rather it is the unique properties of generative AI models that offer the most creative potential. For example, providing different version of a model that have been trained with different sizes of datasets.
    
    \item {\em Prioritize machine agency over human agency:} Increasing controlability and generality hinders a model's ability to assist with creative ideation. A greater degree of machine agency within a collaboration, instead, takes advantage of the unique and unpredictable nature of generative models towards divergent thinking. 

    \item {\em Place constraints on the interaction:} The current iteration of TTI systems are geared towards production, rather than creative ideation. An often cited driver of creative innovation is the use of constraints on the person's process; to push the creator outside of their comfort zone, forcing them to approach problems in a typical manner to overcome the imposed constraints \cite{Harford2016}. In contrast, the generality of the TTI interface necessitates a large degree of oversight from the human collaborator. Moving away from general interfaces for generative AI, towards specified, creative interfaces that exploit machine agency, would result in more fruitful creative ideation.
    
\end{itemize}

\subsubsection{AI for Production}
When using AI for production, the following design considerations are required:

\begin{itemize}
    \item {\em Prioritize human agency over machine agency:} Greater degree of control over the image generation process is required for these systems to become viable tools for creative production. In this case, the balance between agencies should lean towards the human, with controlability of the output as crucial. This could include, for example, the ability to isolate and alter certain elements of an image, a closer adherence to the given prompt, and a greater ability to dictate the style of the output image.
    
    \item {\em Provide domain-specific customization:} Optimising for domain-specific use cases called for by a particular creative practice. For example, the ability to alter the line width of a drawing or to transform a drawing into a stencil (in the case of tattooing), or to variate upon fabric textures (in the case of apparel design).

    \item {\em Integrate with other CSTs:} Integration of generative models into existing software used by creative professionals is key to streamlining generative AI for creative production. We have already seen this kind of integration into creative tools such as the Adobe Suite. 
    
\end{itemize}

Along with the proposed design considerations, we furthermore have gathered a number of insights about the philosophical and ethical implications of developing and integrating generative AI systems. Firstly, we found that all participants conceptualise the system as a creative assistant, or collaborator rather than an artist in it's own right. This relationship emerged naturally, with several participants furthermore expressing the ways in which their personalised model represented an extension of their own aesthetic, their identity, or their  ``brand''.

Participants responded ambivalently to the current ethical controversy surrounding generative AI models. Concerns around copyright and plagiarism, while taken seriously by participants, did not totally discourage their willingness to use a future iteration of a similar system. Furthermore, the particular stance one may take on this issue appeared to reflect their current circumstances within a given creative industry; what one might have to gain or lose from the increasing ubiquity of such a technology. Our analysis hinted that it was perhaps more complex than the current dominant narratives might afford. 

Ambivalence was, in fact, a response we repeatedly observed. We found that the moments in which the model was perceived as `good' (useful output, creatively inspiring, enjoyable to use), came along with it the `bad': perceived threat to ones livelihood, fears around privacy, anxiety around changing role of the human artist. The `better' our participants found the model to be, the more inclined they were to express concern or apprehension about it's development. Similarly, disappointment with the models capabilities was reassuring and reaffirming of the participants role and identity as creators. 

This ambiguity furthermore reiterates the benefit of using the diffractive methodology in evaluating AI systems. If we set out to answer only the question \textit{Is prompt-based generative AI good for creativity?} we would miss the multiple and nuanced interpretations and implications that this technology may present for humanity. Increasingly so, technology can be classified as neither `good' nor `bad', for it does not set out to achieve a singular or specific goal. On the contrary, AI is progressively moving towards generality. A diffractive methodology allowed us to identify the good in the bad, the bad in the good, and ultimately to trace the lines of accountability and influence in the development and application of generative AI models.

\subsection{Limitations and Future Work}
\label{ss:limitations}
We limited the study to one TTI system, Stable Diffusion (V1.4), since the model is open source, and our ability to finetune the model was considered important based on prior literature \cite{Ko2023}. It is also important to clarify that we were not seeking to evaluate the interface or usability aspects of our customised system. We used a basic web interface (Figure \ref{fig:interface}) with a number of controls, and participants were free to experiment with these controls if they wished, however we did not consider the usability aspects of this interface in our analysis. While all participants were comfortable with using traditional creative software, we cannot completely discount how the interface and usability aspects of our system may have impacted on participants' experience. Moreover, alternative TTI systems vary considerably in terms of what imagery they produce given an identical text prompt. Finally, our study only considered the use of TTI systems from the creators perspective. In future studies we will take into account how these systems are understood from an audience perspective.

While our study may be restricted in size, demographic, and diversity, it's worth noting that the artists span a spectrum of experience, from emerging to established professionals. We deliberately selected artists unfamiliar with generative AI to minimize any preconceived biases. Our focused sample size was a strategic choice, enabling us to more closely examine divergences between and within the individual artists. A larger sample size would undoubtedly capture a more comprehensive and extreme set of responses, yet it's precisely our concentrated focus that allowed us to unpack the nuanced intricacies often overlooked in more extensive surveys. As an emerging methodology in HCI research, diffractive analysis is still in its infancy. Future research will further refine and formalise this nascent approach. Despite its preliminary nature, the insights yielded in this study underscore its timeliness, critical relevance, and potential in understanding the impact of AI on the humans.

\section{Conclusion}

The meteoric rise of prompt-based diffusion models mark a profound cultural shift in the space of machine-generated imagery. TTI systems were made accessible to the general public for little to no cost, allowing the immediate generation of detailed and complex imagery. Almost overnight, we were afforded the capacity to produce convincingly `human' visual artwork in the matter of seconds. This profound shift challenged the exclusive role of human as creative agent, and raised concerns around whether it is even necessary to continue to learn and master traditional skills such as illustration, painting, or photography. Moreover, this new generative power also entails new responsibilities: the ethical responsibilities of appropriation and theft of training data\footnote{Open AI recently added a mechanism for websites to be excluded from training (\url{https://platform.openai.com/docs/gptbot}), but this remains an ``opt-out'' rather than ``opt-in'' mechanism and is only for training of ChatGPT.}; the creative agency of AI systems and their yet unknown long-term impact on human culture; the threat of replacing `traditional' creative skills and mediums; and the potential shift of creative economies away from human creators towards large technology companies.

At the time of publishing this research, the initial hype around prompt-based tools has since subsided. The artists in our study responded to the system with skepticism and ambivalence, demonstrating that TTI systems still have a way to go in terms of uptake amongst artists with an established creative practice. 
Our diffractive analysis further identified two modes: \emph{AI for ideation} and \emph{AI for production}. All participants found inadequacies of TTI systems for both use-cases, calling for a different set of design considerations for interface design. 
Nevertheless, the artists were able to envision the integration of a future iteration of image generators in the greater creative practice, whether it be towards expanding their creativity or increasing their efficiency. In this future, they envision the role of AI as a creative assistant, rather than an artist in its own right. Finally, the colonization of AI into traditionally human domains generated ambivalent feelings across the artists; with AI perceived as both a saviour and a threat, often at the same time. 

Our study offers preliminary insight into how artists conceptualise and integrate TTI systems into their existing practice. Yet the reach and ubiquity of these tools herald cultural changes and potential disruptions on a scale and level far beyond the scope of this research. The full ramifications of these shifts, much like the tools themselves, remain the subject of ongoing development.

\begin{acks}
This research was supported by Australian Research Council Grant DP220101223.
\end{acks}

\bibliographystyle{ACM-Reference-Format}
\bibliography{acmart,references,acm_refs}

\appendix

\end{document}